\documentclass[twocolumn,showpacs,prd,floatfix,nofootinbib,showkeys]{revtex4-1}
\usepackage{graphicx}
\usepackage{amssymb}
\usepackage{amsmath}
\usepackage{bm}
\usepackage{natbib}
\newcommand{\p}[1]{(\ref{#1})}
\def\Journal#1#2#3#4{{#1} {\bf #2}, #3 (#4)}

\def\npa{{Nucl. Phys.} A}

\def\plb{{Phys. Lett.}  B}
\def\prl{Phys. Rev. Lett.}
\def\prc{{Phys. Rev.} C}
\def\prd{{Phys. Rev.} D}
\def\aap{Astron. Astrophys.}
\def\apj{Astrophys. J.}

\def\ptp{Prog. Theor. Phys.}
\def\ppnp{Prog. Part. Nucl. Phys.}
\def\jpg{{J. Phys.} G}

\def\jkas{J. Korean Astron. Soc.}

\def\ijmpa{Int. J. Mod. Phys. A}
\def\epja{Eur. Phys. J. A}
\def\epl{Europhys. Lett.}
\def\ptps{Prog. Theor. Phys. Supplement}
\begin{document}
\title[]{Anisotropic pressure in strange quark matter
 in \\ the presence  of a strong nonuniform magnetic field}
\author{A. A.  \surname{Isayev}}
\email{isayev@kipt.kharkov.ua}
  \affiliation{Kharkov
Institute of Physics and Technology, Academicheskaya Street 1,
 Kharkov, 61108, Ukraine }
%

\date{\today}

\begin{abstract}
Thermodynamic properties of strange quark matter (SQM) in a
nonuniform magnetic field are studied within the phenomenological
MIT bag model under the charge neutrality and beta equilibrium
conditions, relevant to the interior of strange quark stars. The
spatial dependence of the magnetic field strength is modeled by the
dependence on the baryon chemical potential in the exponential and
power forms. The total energy density, longitudinal and transverse
pressures in magnetized SQM are found as functions of the baryon
chemical potential. It is clarified that the central magnetic field
strength in a strange quark star is bound from above by the critical
value at which the derivative of the longitudinal pressure with
respect to the baryon chemical potential vanishes first somewhere in
the interior of a star under varying the central field. Above this
upper bound, the instability along the magnetic field is developed
in magnetized SQM. The change in the form of the dependence of the
magnetic field strength on the baryon chemical potential between
the exponential and power ones 
has a nonnegligible effect on the critical magnetic field strength
while the variation of the bag pressure within the absolute
stability window for magnetized SQM has a little effect on
the critical field. 
\end{abstract}



\maketitle

\section{Introduction. Basic Equations}
\label{I}

It was conjectured some time ago that for a certain range of model
QCD-related parameters strange quark matter (SQM), consisting of
deconfined $u$, $d$ and $s$ quarks, can be the true ground state of
matter~\cite{AB,W,FJ}. In that case, at zero external pressure and
temperature, the energy per baryon of SQM is less than that for the
most stable $^{56}$Fe nucleus. If this hypothesis holds true, then
the formation of strange quark stars, composed of SQM and self-bound
by strong interactions, is possible~\cite{I,AFO,HZS,FW}. This
conjecture got recently support from the  observations of massive
compact stars with $M\sim2M_\odot$ and the indications that some
neutron stars may be very compact (with the radii smaller than
10~km). While the former implies that the equation of state (EoS) of
strongly interacting matter should be stiff, the latter requires the
soft EoS. A possible explanation of these contradictory requirements
could be the existence of two separate families of compact stars:
quark stars which can be very massive, according to the perturbative
QCD calculations, and hadronic stars which can be very
compact~\cite{PRD14D}.

 Another important peculiarity, related to compact stars, is that
they can possess strong magnetic fields. For example, for magnetars
--- strongly magnetized neutron stars~\cite{DT}, the magnetic field
strength can reach  values of about $10^{14}$–-$10^{15}$~G at the
surface~\cite{TD,IShS}, and can be even larger, up to $10^{19}$~G,
in the core of a star~\cite{CBP,BCP}. Usually, such estimates of the
possible interior magnetic field strengths are based on the 
virial  theorem~\cite{ApJ53Chandrasekhar} while general
relativistic 
calculations, based on the Einstein--Maxwell equations, lead to the
more modest estimate \hbox{$H\lesssim (1$--$3)$ $\times
10^{18}$}~G~\cite{PLB02Broderick}. Such strong magnetic fields can
result in the large pulsar kick velocities because of the asymmetric
neutrino emission in direct Urca processes in the dense core of a
magnetized compact star~\cite{ARDM}. The mechanism responsible for
generation of strong magnetic fields of magnetars is still to be
clarified, and, among other possibilities, this can be due to the
turbulent dynamo amplification mechanism in a star with the rapidly
rotating core~\cite{TD}, or because of the spontaneous ordering of
nucleon~\cite{IY,PRC05I,PRC06I,PRC07I}, or quark~\cite{TT} spins in
the dense matter inside a compact star.


Strong magnetic fields can have significant impact on thermodynamic
properties of strongly interacting matter in the dense interior of a
compact star~\cite{BPL,RPPP,IY4,IY10,PRD12Wen,PRD15Chu}.
 In
particular, because of the breaking of the rotational symmetry, the
pressure becomes essentially anisotropic in strongly magnetized
matter~\cite{Kh,FIKPS,PRD11Paulucci,IY_PRC11,IY_PLB12,JPG13IY}. The
longitudinal pressure $p_l$ (along the magnetic field direction)
gets the negative contribution from the magnetic field given by the
Maxwell term $\frac{H^2}{8\pi}$. Under increasing the magnetic field
strength, the longitudinal pressure decreases and, eventually,
becomes negative, resulting in the appearance of the longitudinal
instability in a strongly magnetized matter. For the uniform magnetic
field, the onset of the longitudinal instability corresponds to the
critical magnetic field strength, at which the longitudinal pressure
vanishes. The estimates show that the critical magnetic field has
the upper bound of about $10^{19}$~G for quark
matter~\cite{FIKPS,PRD11Paulucci,JPG13IY}, neutron
matter~\cite{IY_PRC11,IY_PLB12} and strange baryonic
matter~\cite{NPA13SMS}.  For strange quark stars, self-bound by
strong interactions, the condition  of absolute stability of
magnetized SQM, with account of the pressure anisotropy, sets the
constraint on the allowable magnetic field strength \hbox{$H\lesssim
(1$--$3)$ $\times 10^{18}$}~G~\cite{IJMPA14I,PRC15I}. For hybrid
stars, based on the energy conservation arguments, the possible
magnetic field strength in the quark core is estimated as
$H\sim10^{20}$~G~\cite{FIKPS}. For a nonuniform magnetic field, with
allowance for the inhomogeneous mass distribution, the application
of the virial theorem gives the estimate for the central field in a
neutron star $H\sim10^{19}$~G~\cite{FIKPS}.

In this research,  we  study thermodynamic properties of SQM in a
nonuniform magnetic field, taking into account that in strange quark
stars the magnetic field strength can change by several orders of
magnitude from the core to the surface of a star. The spatial
dependence of the magnetic field strength is modeled by its
dependence on the baryon chemical potential $\mu_B$.  As will be
shown in this study,  the longitudinal instability in  a nonuniform
magnetic field is associated with the appearance of the negative
derivative $p_{\,l}^{\,\prime}(\mu_B)<0$, unlike to the case of an
uniform magnetic field where the longitudinal instability occurs at
$p_{\,l}<0$.

As a theoretical framework to study strongly magnetized SQM, we will
utilize the MIT bag model. The details of a theoretical formalism
are presented in
Ref.~\cite{JPG13IY}. 
The domain of absolute stability of magnetized SQM within the MIT
bag model with account of the effects of the pressure anisotropy was
determined in~\cite{IJMPA14I,PRC15I}.
 To mimic the spatial dependence of the magnetic
field, we will parametrize  the magnetic field strength in terms of
the baryon chemical potential $\mu_B$ in the exponential
form~\cite{EPJA12Dexheimer,JPG14Dexheimer,PRD15Carignano}:
\begin{align}
     H(\mu_B)= H_{s}+H_{cen} 
\Bigl(1-e^{-\beta\bigl(\frac{\mu_B-\mu_{B0}}{\mu_{B0}}\bigr)^\gamma}\Bigr)
\label{H_muB}.
\end{align}
Here $\mu_{B0}$ and $H_s$ are the baryon chemical potential and the
magnetic field strength at the surface of a strange quark star,
respectively. In Eq.\p{H_muB}, the quantity $H_{cen}$ is given by
$H_{cen}\approx H(\mu_B\gg\mu_{B0})$, assuming that $H_{cen}\gg H_s$; 
 $\beta$ and $\gamma$ are the model parameters. Also, in numerical
 calculations we will adopt the power parametrization:
\begin{align}
     H(\mu_B)= H_{s}+H_{cen} 
\Bigl[1-\Bigl(\frac{\mu_{B}^c-\mu_{B}}{\mu_{B}^c-\mu_{B0}}\Bigr)^\alpha\Bigr],\label{pow}
\end{align}
where $\mu_{B}^c$ is the baryon chemical potential in the center of
a star, $\alpha$ is the model parameter. One can see that
$H(\mu_{B0})=H_{s}$ and $H(\mu_{B}^c)\approx H_{cen}$. Further we
consider bare strange quark stars (without a thin layer of nuclear
matter above the quark surface). In order to determine the baryon
chemical potential $\mu_{B0}$ at the surface, we will use the
conditions of charge neutrality
\begin{align}2\varrho_u-\varrho_d-\varrho_s-3\varrho_{e}=0,\label{cnc}\end{align}
and chemical equilibrium with respect to weak processes in SQM:
\begin{align}\mu_d&=\mu_u+\mu_{e},\label{mud}\\
\mu_d&=\mu_s, \label{ds}
\end{align}
where $\varrho_i$ and $\mu_i$  are the number density and chemical
potential for fermions of $i$th species ($i=u,d,s,e$). Further
 we assume, analogously to Ref.~\cite{PRD14Chu}, a spherically
symmetric radial distribution of the magnetic field inside a star.
Then the longitudinal pressure (along the magnetic field direction)
should vanish at the surface of a star:
\begin{align}
p_{\,l}&=-\sum_i\Omega_i-\frac{H^2}{8\pi}-B
=0.\label{p_l}\end{align} Here $\Omega_i$ is the thermodynamic
potential  for free relativistic fermions of $i$th species in a
magnetic field~\cite{PRD12Wen,JPG13IY}:
\begin{align}\label{omegai0}\Omega_i&=-\frac{|q_i|g_i H}{4\pi^2}\sum_{\nu=0}^{\nu_{\mathrm{max}}^i}
(2-\delta_{\nu,0}) \\ &\quad \times \biggl\{\mu_i k^i_{F,\nu}-\bar
m_{i,\nu}^2\ln\biggl|\frac{k^i_{F,\nu}+\mu_i}{\bar
m_{i,\nu}}\biggr|\biggr\},\nonumber
\end{align}
and
\begin{align}\bar
m_{i,\nu}=\sqrt{m_i^2+2\nu|q_i|H},\quad
k^i_{F,\nu}=\sqrt{\mu_i^2-\bar m_{i,\nu}^2}\,.
\end{align}
In Eq.~\p{omegai0}, summation on Landau levels  runs up to
$$\nu_{\mathrm{max}}^i=I\bigl[\frac{\mu_i^2-m_i^2}{2|q_i|H}\bigr],$$
$I[...]$ being an integer part of the argument, the factor
$(2-\delta_{\nu,0})$ takes into account the spin degeneracy of
Landau levels, and $g_i$ is the remaining degeneracy factor ($g_i=3$
if $i=u,d,s$ and $g_i=1$ if $i=e$).

 The chemical potentials of
quarks and electrons at the surface of a quark star can be
determined from Eqs.~\p{cnc}--\p{p_l} with account of the
relationship between the particle number densities
$\varrho_i=-\bigl(\frac{\partial\Omega_i}{\partial \mu_i}\bigl)_H$
and respective chemical potentials $\mu_i$:
\begin{align}\varrho_i=\frac{|q_i|g_i
H}{2\pi^2}\sum_{\nu=0}^{\nu_{\mathrm{max}}^i}
(2-\delta_{\nu,0})k^i_{F,\nu}. \label{rhoi}
\end{align}
    Since
\begin{align}\mu_{B}=\mu_{u}+\mu_{d}+\mu_{s}, \label{muB}
\end{align}
 then one can   find  the
 baryon chemical potential $\mu_{B0}$ at the surface of a star. In numerical
calculations, we will use the model parameters $\beta=45$ and
$\gamma=3$ for the exponential parametrization, $\alpha=\frac{3}{2}$
for the power parametrization, and $H_s=10^{15}$~G.
 The bag pressure is set
$B=74$~MeV/fm$^3$, which is slightly smaller than the upper bound
$B_u\simeq75$~MeV/fm$^3$ from the absolute stability window for the
quark current  masses $m_u=m_d=5$~MeV, and
$m_s=150$~MeV~\cite{IJMPA14I}. Then one can numerically determine
the baryon chemical potential $\mu_{B0}\approx927.4$~MeV.

\section{Numerical results and discussion }

\begin{figure*}
\begin{center}
 \includegraphics[height=6.9cm]{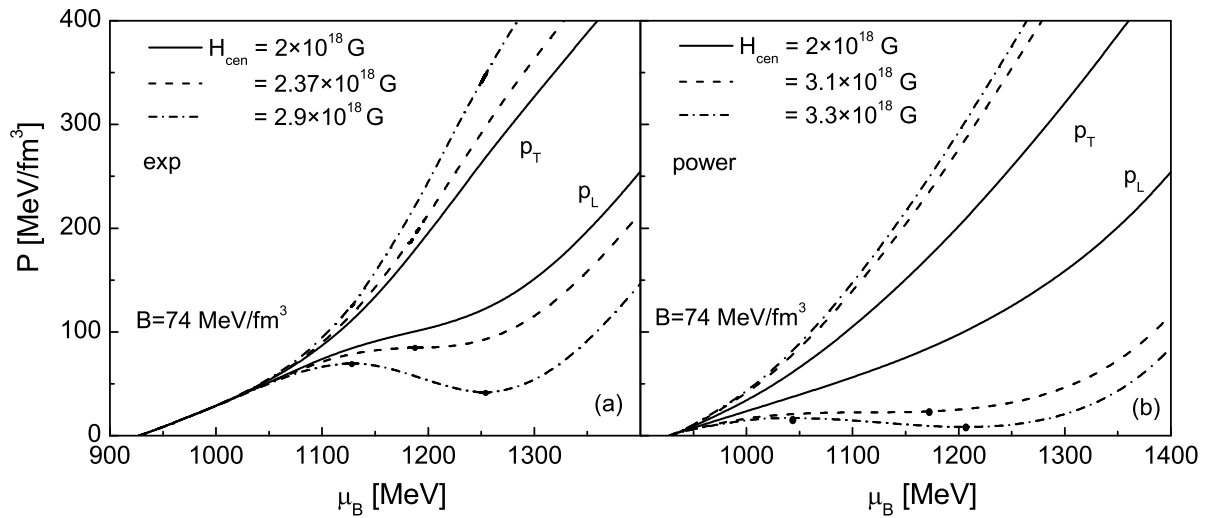}
\end{center}
\vspace{-0ex} \caption{Transverse $p_t$ 
and
longitudinal $p_l$ 
pressures in magnetized SQM as functions of the baryon chemical
potential, corresponding to: (a) the exponential
parametrization~\p{H_muB} with $\beta=45,\gamma=3$, and (b) the
power parametrization~\p{pow} with $\alpha=\frac{3}{2}$,
$\mu_B^c=1400$~MeV for $H_s=10^{15}$~G and variable central field
$H_{cen}$. The full dots correspond to the points where
$p_{\,l}^{\,\prime}(\mu_B)=0$.} \label{fig1} \vspace{-0ex}
\end{figure*}

In the MIT bag model, the total energy density $E$ and the
transverse pressure $p_{\,t}$ in magnetized SQM read
\begin{align}
E&=\sum_i \bigl(\Omega_i+\mu_i\varrho_i\bigr)+\frac{H^2}{8\pi}+B,\label{E}\\
p_{\,t}&=-\sum_i\Omega_i-HM+\frac{H^2}{8\pi}-B, \label{p_t}
\end{align}
where $M=-\sum_i \bigl(\frac{\partial\Omega_i}{\partial
H}\bigr)_{\mu_i}$ is the magnetization of the system. In order to
study the impact of a strong nonuniform magnetic field, parametrized
by Eq.~\p{H_muB}, or by Eq.~\p{pow}, on the anisotropic pressure and
the equation of state (EoS) of the system, we will fix the  baryon
chemical potential in the center of a strange quark star,
$\mu_B^c=1400$~MeV (that corresponds to the baryon number density
$\varrho_B^c$ of about eight times nuclear saturation density ---
densities of such magnitude are expected to occur in the center of
strange quark
stars~\cite{PRD11Paulucci,EPJA12Dexheimer,JPG14Dexheimer}), and will
vary the central magnetic field strength $H_{cen}$.

Fig.~\ref{fig1} shows the dependence of the transverse $p_{\,t}$ and
longitudinal $p_{\,l}$   pressures in the system on the baryon
chemical potential $\mu_B$ for several values of the central
magnetic field strength $H_{cen}$. Let us discuss, first, the case
of the exponential parametrization of the magnetic field strength,
represented in Fig.~\ref{fig1}a. It is seen that, under increasing
the central field $H_{cen}$, the transverse pressure $p_{\,t}$
increases while the longitudinal pressure $p_{\,l}$ decreases. Also,
the transverse pressure $p_{\,t}$ always remains the increasing
function of the baryon chemical potential $\mu_B$ while the
dependence of the longitudinal pressure $p_{\,l}$ on $\mu_B$ can be
different. At not too strong central fields (e.g., at
$H_{cen}=2\cdot10^{18}$~G), the longitudinal pressure $p_{\,l}$
remains the increasing function of $\mu_B$. However, with the
increase of $H_{cen}$, the curve $p_{\,l}(\mu_B)$ bends down in its
middle  part, and there exists such central field $H_{cen}$, at
which the derivative $p_{\,l}^{\,\prime}(\mu_B)$ vanishes first
somewhere in the interior of a strange quark star. For a given set
of the model parameters, this happens for
$H_{cen}\approx2.37\cdot10^{18}$~G at $\mu_B\approx1188.4$~MeV (the
corresponding point on the curve is marked by the full dot). Under
further increasing the central field $H_{cen}$, there appears the
part on the curve $p_{\,l}(\mu_B)$, characterized by
$p_{\,l}^{\,\prime}(\mu_B)<0$ (e.g.,  at
$H_{cen}\approx2.9\cdot10^{18}$~G, this part of the curve on the
figure is contained between two full dots). This contradicts  the
thermodynamic  constraint $p_{\,l}^{\,\prime}(\mu_B)>0$. Hence, such
states of magnetized SQM are unstable, and instability is developed
along the magnetic field direction. The strength of the central
field $H_{cen}\approx2.37\cdot10^{18}$~G, at which the derivative
$p_{\,l}^{\,\prime}(\mu_B)$ vanishes first, is the critical field
for the onset of the longitudinal instability. This value represents
the upper bound on the central magnetic field strength in a strange
quark star.

Concerning the criterion $p_{\,l}^{\,\prime}(\mu_B)<0$ for the
appearance of the longitudinal instability, it is important to note
that, according to rigorous microscopic
derivations~\cite{FIKPS,PRD12Strickland}, the total parallel
pressure $p_{\,l}$ contains both matter and field contributions (cf.
Eq.~\p{p_l}). While the  derivative of the matter part of  the
longitudinal pressure with respect to the baryon chemical potential
is positive, the magnetic field contributes negatively to the
derivative $p_{\,l}^{\,\prime}(\mu_B)$, and, at a strong enough
central field, the field contribution overcomes the matter
contribution, making the derivative $p_{\,l}^{\,\prime}(\mu_B)$
negative.

\begin{figure*}
\begin{center}
 \includegraphics[height=6.9cm]{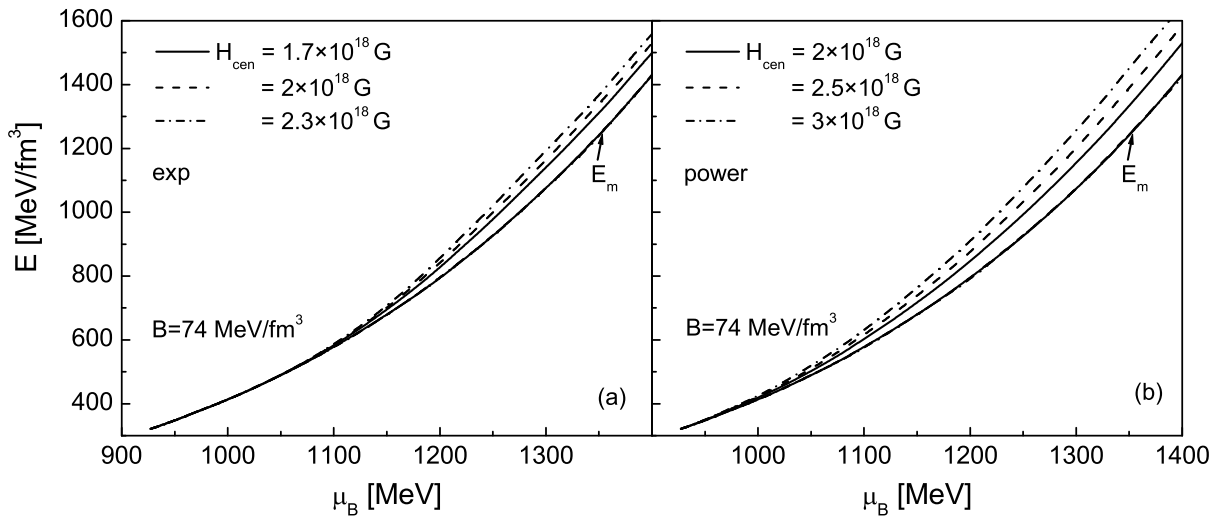}
\end{center}
\vspace{-0ex} \caption{Same as in Fig.~\ref{fig1}, but for the
energy density $E$  and the matter energy density $E_m\equiv
E-\frac{H^2}{8\pi}$ (the most lower curve) of magnetized SQM.}
\label{fig2} \vspace{-0ex}
\end{figure*}

For the power parametrization~\p{pow} of the magnetic field strength
(cf. Fig.~\ref{fig1}b), the behavior of the curves $p_{\,t}(\mu_B),
p_{\,l}(\mu_B)$ under varying the central field $H_{cen}$ is
qualitatively similar to that for the exponential
parametrization~\p{H_muB}. The derivative
$p_{\,l}^{\,\prime}(\mu_B)$ vanishes first for
$H_{cen}\approx3.1\cdot10^{18}$~G at $\mu_B\approx1172.1$~MeV, and,
hence, the change in the form of the parametrization $H(\mu_B)$ from
exponential to the power one has the nonnegligible effect on the
critical magnetic field strength.

We repeated also the calculations for the bag pressure
$B=58$~MeV/fm$^3$ which is slightly above the lower bound
$B_l\simeq57$~MeV/fm$^3$ from the absolute stability
window~\cite{IJMPA14I}, but this variation of the bag pressure has a
little effect on the results, for example, for the power
parametrization the critical magnetic field strength increases till
$H_{cen}\approx3.17\cdot10^{18}$~G.

Note that for an uniform magnetic field the longitudinal instability
in  magnetized matter is associated with the appearance of the
negative longitudinal pressure
$p_l<0$~\cite{Kh,FIKPS,PRD11Paulucci,IY_PRC11,IY_PLB12,JPG13IY}.
Since vanishing of the derivative $p_{\,l}^{\,\prime}(\mu_B)$ occurs
at smaller central magnetic field strength $H_{cen}$ than vanishing
of $p_{\,l}$, for a nonuniform magnetic field, parametrized by
Eq.~\p{H_muB}, or by Eq.~\p{pow}, the corresponding criterion is the
occurrence of the negative derivative $p_{\,l}^{\,\prime}(\mu_B)<0$.
The last criterion for the appearance of the longitudinal
instability in a nonuniform magnetic field sets a stronger
constraint on the upper bound of the central magnetic field in
strongly
magnetized strange quark star than the criterion 
$p_{\,l}<0$.

Fig.~\ref{fig2} shows the  energy density $E$ of magnetized SQM and
its matter part $E_m\equiv E-\frac{H^2}{8\pi}$ (without the pure
magnetic field contribution)  as functions of the baryon chemical
potential. With increasing the central magnetic field strength, the
energy density $E$ increases  while the  matter part $E_m$ remains
practically unchanged. In particular, the curves for  $E_m$ are
almost indistinguishable for the different values of the central
magnetic field $H_{cen}$, used in calculations, and look as one
curve. This figure allows to estimate the relative role of the
matter $E_m$ and magnetic field $\frac{H^2}{8\pi}$ contributions to
the total energy density $E$. It is seen that the matter part
dominates over the field part at such baryon chemical potentials and
magnetic field strengths $H_{cen}$ for both the exponential and
power parametrizations of the magnetic field strength.

\begin{figure*}[tb]
\begin{center}
 \includegraphics[height=6.9cm]{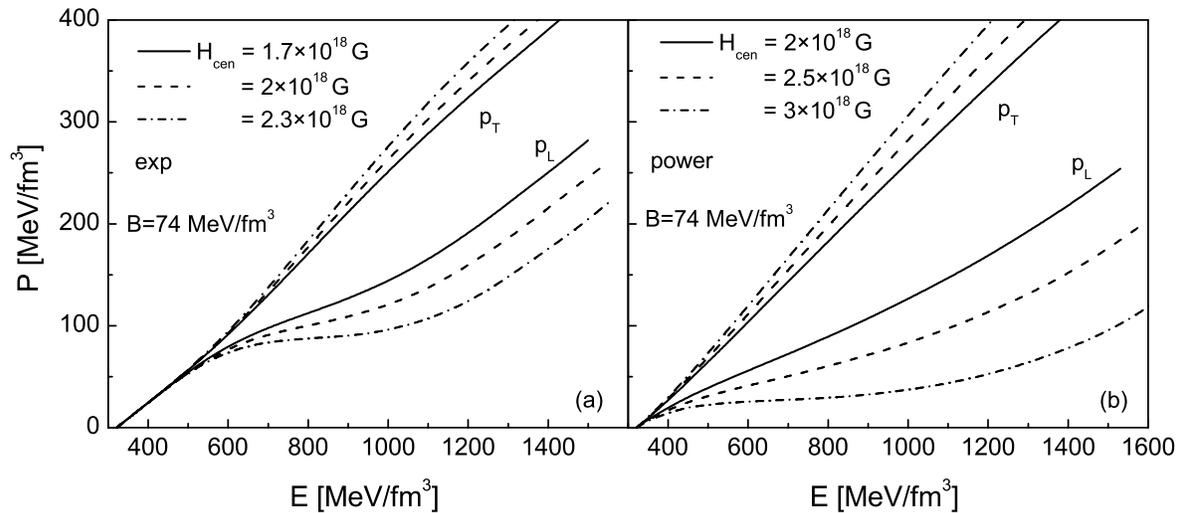}
\end{center}
\vspace{-1ex} \caption{The transverse $p_t$ and longitudinal $p_l$
pressures in magnetized SQM as functions of the total energy density
$E$, corresponding to: (a) the exponential
parametrization~\p{H_muB}, and (b) the power parametrization~\p{pow}
of the magnetic field strength. 
}
\label{fig3} \vspace{-0ex}
\end{figure*}

In   strong magnetic fields  the total pressure in magnetized SQM
becomes essentially anisotropic. Therefore, EoS of the system
becomes also highly anisotropic. Fig.~\ref{fig3}, showing the
dependence of the transverse $p_t$  and longitudinal $p_l$ pressures
on the energy density $E$ of magnetized SQM, explicitly demonstrates
this moment. In the given cases, when  the values of the central
field are smaller than the critical field for the appearance of the
longitudinal instability, the pressures $p_t$, $p_l$, and the energy
density $E$ are the increasing functions of the baryon chemical
potential $\mu_B$. Hence, after excluding $\mu_B$, one gets the
anisotropic EoS in the form of two distinct increasing functions
$p_t(E)$ and $p_l(E)$.

In conclusion, we have considered the impact of a strong magnetic
field  on thermodynamic properties of SQM at zero temperature under
conditions relevant to the interior of  magnetized strange quark
stars. The spatial dependence of the magnetic field strength is
modeled by the dependence on the baryon chemical potential in
the exponential and power forms. The total energy density $E$,
transverse $p_t$
 and  longitudinal $p_l$ pressures in magnetized SQM have been calculated as functions of the  baryon
chemical potential. Also,  the highly anisotropic EoS has been
determined in the form of $p_t(E)$ and $p_l(E)$ dependences. It has
been clarified that the central magnetic field in a strange quark
star is bound from above by the critical value, at which  the
derivative of the longitudinal pressure $p_{\,l}^{\,\prime}(\mu_B)$
vanishes first somewhere in the interior of a star under varying the
central field. Above this upper bound, the instability along the
magnetic field direction is developed in magnetized SQM. The change
in the form of the dependence $H(\mu_B)$ between the exponential and
power ones leads to the noticeable quantitative differences, in
particular, it has the non-negligible effect on the critical magnetic
field strength. While the variation of the bag pressure within the
absolute stability window for magnetized SQM has a little effect on
the results, in particular, the critical field remains almost
unaltered under such a change.

Based on the criterion of the longitudinal instability
$p_{\,l}^{\,\prime}(\mu_B)<0$, the possible central magnetic field
strength \hbox{$H_{cen}\lesssim (2$--$3)$ $\times 10^{18}$}~G has
been estimated to be more than three orders of magnitude larger than
the surface field. In some of the previous calculations, based on
the Einstein--Maxwell equations, the central magnetic field was
estimated to be only five times larger than the surface
value~\cite{ApJ54Ferraro}. Nevertheless, as discussed in
Ref.~\cite{PLB02Broderick}, where solution of the Einstein--Maxwell
equations gives the estimate on the possible interior magnetic field
\hbox{$H\lesssim (1$--$3)$ $\times 10^{18}$}~G, other choices of the
nonuniform current function, or the relaxation of
the condition of axial symmetry of magnetic field distribution, which can influence the shape of a star, 
could lead to even stronger interior magnetic fields.

In this research, all consideration has been done within the
phenomenological MIT bag model, which is quite popular and
frequently used in various astrophysical applications (just some of
recent references include, e.g.,
\cite{PRD12Wen,EPJA11Liu,EPL13Torres,JPG14Dexheimer,arxiv17Cardoso}).
Despite its relative simplicity, it allows to qualitatively describe
the appearance of the longitudinal instability in strongly
magnetized SQM and to get the correct order of magnitude of the
upper bound on the  magnetic field strength in strange quark stars.
The MIT bag model establishes the baseline for more advanced
calculations and further improvement of the obtained estimates is
possible with more elaborated models.

It is worthy to note that the proposed mechanism for the appearance
of the longitudinal instability in magnetized matter in a nonuniform
magnetic field parametrized in terms of the baryon chemical
potential is universal and does not depend on the specific type of a
compact star whether it is a quark star, or a neutron star, or a
hybrid star. The specific type of a compact star will be reflected
in the underlying model for the EoS of matter in the interior of
compact stellar object, depending on whether it is a quark phase, or
a hadronic phase in the given inner region of a star. Inevitably,
the longitudinal instability will occur in a strong enough magnetic
field as soon as the derivative $p_{\,l}^{\,\prime}(\mu_B)$ becomes
negative in the field beyond the critical one.

The formulation of the problem  in terms of the energy density $E$
as an independent variable would be the other possible way to
consider the longitudinal instability in a strong nonuniform
magnetic field. This would lead to the criterion of the longitudinal
instability in the form $p_{\,l}^{\,\prime}(E)<0$. Under such an
approach, it would be consistent to parametrize the magnetic field
strength $H$ in terms of the energy density $E$ as well, $H=H(E)$.
Nevertheless, taking into account the possible applications of the
criterion of the longitudinal instability to other types of compact
stars, such as, e.g., hybrid stars, the most flexible way  to tackle
the problem is to formulate it in terms of the baryon chemical
potential. If the energy density were  used as the independent
variable, then, because the energy density is
discontinuous 
at the phase boundary
  of a first order quark-hadron phase transition,
the magnetic field strength $H(E)$ would  experience the unphysical
jump across the phase transition boundary, which is missing for the
parametrization $H(\mu_B)$.

Note that, although the presence of a magnetic field leads to the
appearance of the local pressure anisotropy, magnetized strange
quark stars, considered in this study, are spherically symmetric
because of the radial distribution of the magnetic field inside a
star. There can be other sources of the local  pressure anisotropy,
like superfluid states with the finite orbital momentum of Cooper
pairs~\cite{PTP93Takatsuka,PRC98Baldo,PRC02IR,NPA03Zverev,EPL08Zuo},
or  finite superfluid momentum~\cite{PRC02I,RMP04Casalbuoni}, which,
nevertheless, lead to a spherically symmetric star. It would be of
interest to extend this research by incorporating the effects of the
pressure anisotropy within the framework of general relativity.


\section*{References}
\begin{thebibliography}{99}

\bibitem{AB}A. Bodmer, \Journal{\prd}{4}{1601}{1971}. 

\bibitem{W} E. Witten, \Journal{\prd}{30}{272}{1984}. 

\bibitem{FJ} E. Farhi and R. L. Jaffe, \Journal{\prd}{30}{2379}{1984}. 


\bibitem{I} N. Itoh, \Journal{\ptp}{44}{291}{1970}. 
\bibitem{AFO} C. Alcock, E. Farhi, and A. V. Olinto, \Journal{\apj}{310}{261}{1986}. 

\bibitem{HZS} P. Haensel, J. Zdunik, and R. Schaeffer, \Journal{\aap}{160}{121}{1986}. 

\bibitem{FW} F. Weber, \Journal{\ppnp}{54}{193}{2005}.


\bibitem{PRD14D} A. Drago, A. Lavagno, and G. Pagliara, \Journal{\prd}{89}{043014}{2014}.



\bibitem{DT} R.C. Duncan, and C. Thompson,  \Journal{\apj}{392}{L9}{1992}. 

\bibitem{TD} C. Thompson, and R.C. Duncan,  \Journal{\apj}{473}{322}{1996}. 

\bibitem{IShS}A. I. Ibrahim, S. Safi-Harb, J. H. Swank, W. Parke, and S. Zane,
\Journal{\apj}{574}{L51}{2002}.

\bibitem{CBP} S. Chakrabarty, D. Bandyopadhyay, and S. Pal, \Journal{\prl}{78}{2898}{1997}.

\bibitem{BCP}  D. Bandyopadhyay, S. Chakrabarty,  and S. Pal, \Journal{\prl}{79}{2176}{1997}. 

\bibitem{ApJ53Chandrasekhar} S. Chandrasekhar, and E. Fermi, \Journal{\apj}{118}{116}{1953}.

\bibitem{PLB02Broderick} A. Broderick, M. Prakash, and J. M. Lattimer, \Journal{\plb}{531}
{167}{2002}.

\bibitem{ARDM}
 S. P. Adhya, P. K. Roy and A.
K. Dutt-Mazumder, \Journal{\jpg}{41}{025201}{2014}.

\bibitem{IY}  A.A.  Isayev, and  J. Yang,   \Journal{\prc}{69}{025801}{2004}. 

\bibitem{PRC05I}
A.A. Isayev,  \Journal{\prc}{72}{014313}{2005}.


\bibitem{PRC06I}
A.A. Isayev,  \Journal{\prc}{74}{057301}{2006}. 

\bibitem{PRC07I}
A.A. Isayev,  \Journal{\prc}{76}{047305}{2007}.

\bibitem{TT} T. Tatsumi, \Journal{\plb}{489}{280}{2000}. 

\bibitem{BPL} A. Broderick, M. Prakash, and J. M. Lattimer, \Journal{\apj}{537}{351}{2000}. 
\bibitem{RPPP} A. Rabhi, H. Pais, P.K. Panda,  and C. Provid\^encia, \Journal{\jpg}{36}{115204}{2009}.
\bibitem{IY4}  A.A.  Isayev, and  J. Yang,   \Journal{\prc}{80}{065801}{2009}. 

\bibitem{IY10} A.A.  Isayev, and  J. Yang, \Journal{\jkas}{43}{161}{2010}.

\bibitem{PRD12Wen} X.J. Wen, S.Z. Su, D.H. Yang, and G.X. Peng,   \Journal{\prd}{86}{034006}{2012}. 


\bibitem{PRD15Chu}
P. C. Chu, X. Wang, L. W. Chen, and M. Huang, 
\Journal{\prd}{91}{023003}{2015}.


\bibitem{Kh} V. R. Khalilov, \Journal{\prd}{65}{056001}{2002}. 


\bibitem{FIKPS} E. J. Ferrer, V. de la Incera, J. P. Keith, I. Portillo, and P. L. Springsteen, \Journal{\prc}{82}{065802}{2010}.
\bibitem{PRD11Paulucci} L. Paulucci, E. J. Ferrer,  V. de la Incera, and J. E. Horvath,
\Journal{\prd}{83}{043009}{2011}.

\bibitem{IY_PRC11}  A.A. Isayev, and J. Yang, \Journal{\prc}{84}{065802}{2011}. 

\bibitem{IY_PLB12}  A.A. Isayev, and J. Yang, \Journal{\plb}{707}{163}{2012}. 

\bibitem{JPG13IY} A.A.  Isayev, and  J. Yang, \Journal{\jpg}{40}{035105}{2013}. 

\bibitem{NPA13SMS}M. Sinha, B. Mukhopadhyay,  and A. Sedrakian, \Journal{\npa}{898}{43}{2013}.

\bibitem{IJMPA14I} A.A.  Isayev, \Journal{\ijmpa}{29}{1450173}{2014}.
\bibitem{PRC15I} A.A.  Isayev, \Journal{\prc}{91}{015208}{2015}.
\bibitem{EPJA12Dexheimer} V. Dexheimer, R. Negreiros, and S. Schramm, \Journal{\epja}{48}{189}{2012}.
\bibitem{JPG14Dexheimer} 
V. Dexheimer, D. P. Menezes, and M. Strickland, \Journal{\jpg}{41}{015203}{2014}. 
\bibitem{PRD15Carignano} S. Carignano, E. J. Ferrer, V. de la Incera, and L.~Paulucci,
\Journal{\prd}{92}{105018}{2015}.
\bibitem{PRD14Chu} P.C. Chu, L.W. Chen, and X. Wang, \Journal{\prd}{90}{063013}{2014}.

\bibitem{PRD12Strickland} M. Strickland, V. Dexheimer, and D. P. Menezes, \Journal{\prd}{86}{125032}{2012}.

\bibitem{ApJ54Ferraro} V. C. A. Ferraro,
\Journal{\apj}{119}{407}{1954}.

\bibitem{EPJA11Liu} B. Liu, M. Di Toro, G.Y. Shao, V.Greco, C.W. Shen, and Z.H. Li, \Journal{\epja}{47}{104}{2011}.
\bibitem{EPL13Torres}J. R. Torres, and D. P. Menezes, \Journal{\epl}{101}{42003}{2013}.
\bibitem{arxiv17Cardoso} P. H. G. Cardoso, T. N. da Silva, A. Deppman, and
D.~P.~Menezes, \Journal{\epja}{53}{191}{2017}. 

\bibitem{PTP93Takatsuka}
T. Takatsuka and R. Tamagaki, \Journal{\ptps}{112}{27}{1993}.
 \bibitem{PRC98Baldo}
M. Baldo, O. Elgaroy, L. Engvik, M. Hjorth-Jensen, and H.-J. Schulze,
\Journal{\prc}{58}{1921}{1998}.
\bibitem{PRC02IR} A.A.  Isayev and G. R\"opke, \Journal{\prc}{66}{034315}{2002}.
\bibitem{NPA03Zverev} M. V. Zverev, J. W. Clark,  and V. A. Khodel, \Journal{\npa}{720}{20}{2003}.
\bibitem{EPL08Zuo} W. Zuo, A. J. Mi, C. X. Cui, and U. Lombardo , \Journal{\epl}{84}{32001}{2008}.

\bibitem{PRC02I} A.A.  Isayev, \Journal{\prc}{65}{031302}{2002}.
\bibitem{RMP04Casalbuoni} R. Casalbuoni and G. Nardulli, \Journal{\rmp}{76}{263}{2004}.


\end{thebibliography}

\end{document}